\documentclass{emulateapj}

\usepackage{hyperref, color}

\newcommand{\etal}{et~al.}
\newcommand{\Msun}{M$_{\odot}$}

\newcommand{\Msol}{M$_{\odot}$}

\newcommand{\kms}{\hbox{km~s$^{-1}$}}

\newcommand{\ArII}{\hbox{{\rm Ar}\kern 0.1em{\sc ii}}}
\newcommand{\ArIII}{\hbox{{\rm Ar}\kern 0.1em{\sc iii}}}
\newcommand{\CIV}{\hbox{{\rm C}\kern 0.1em{\sc iv}}}
\newcommand{\HI}{\hbox{{\rm H}\kern 0.1em{\sc i}}}
\newcommand{\HII}{\hbox{{\rm H}\kern 0.1em{\sc ii}}}
\newcommand{\HeI}{\hbox{{\rm He}\kern 0.1em{\sc i}}}
\newcommand{\HeII}{\hbox{{\rm He}\kern 0.1em{\sc ii}}}
\newcommand{\NII}{\hbox{{\rm N}\kern 0.1em{\sc ii}}}
\newcommand{\OI}{\hbox{{\rm O}\kern 0.1em{\sc i}}}
\newcommand{\OII}{\hbox{{\rm O}\kern 0.1em{\sc ii}}}
\newcommand{\OIII}{\hbox{{\rm O}\kern 0.1em{\sc iii}}}
\newcommand{\OIIlong}{{\rm O}\kern 0.1em{\sc ii}~$\lambda 3727$} 
\newcommand{\FeII}{\hbox{{\rm Fe}\kern 0.1em{\sc ii}}}
\newcommand{\NeII}{\hbox{{\rm Ne}\kern 0.1em{\sc ii}}}
\newcommand{\NeIII}{\hbox{{\rm Ne}\kern 0.1em{\sc iii}}}
\newcommand{\NeV}{\hbox{{\rm Ne}\kern 0.1em{\sc v}}}
\newcommand{\SII}{\hbox{{\rm S}\kern 0.1em{\sc ii}}}
\newcommand{\SIII}{\hbox{{\rm S}\kern 0.1em{\sc iii}}}
\newcommand{\SIV}{\hbox{{\rm S}\kern 0.1em{\sc iv}}}
\newcommand{\SiIV}{\hbox{{\rm Si}\kern 0.1em{\sc iv}}}
\newcommand{\MgII}{\hbox{{\rm Mg}\kern 0.1em{\sc ii}}}
\newcommand{\Halpha}{\hbox{{\rm H}\kern 0.1em$\alpha$}}
\newcommand{\Hbeta}{\hbox{{\rm H}\kern 0.1em$\beta$}}
\newcommand{\Heopta}{\hbox{{\rm He}\kern 0.1em{\sc i}}~$6678$}
\newcommand{\Heoptb}{\hbox{{\rm He}\kern 0.1em{\sc i}}~$5876$}
\newcommand{\Heoptc}{\hbox{{\rm He}\kern 0.1em{\sc i}}~$4471$}
\newcommand{\Brgam}{\hbox{{\rm Br}\kern 0.1em$\gamma$}}
\newcommand{\Brten}{\hbox{{\rm Br}\kern 0.1em$10$}}
\newcommand{\Breleven}{\hbox{{\rm Br}\kern 0.1em$11$}}
\newcommand{\HeIh}{\hbox{{\rm He}\kern 0.1em{\sc i}}~$1.7$~{\micron}}
\newcommand{\HeIk}{\hbox{{\rm He}\kern 0.1em{\sc i}}~$2.06$~{\micron}}
\newcommand{\squishlist}{
   \begin{list}{$\bullet$}
    { \setlength{\itemsep}{0pt}      \setlength{\parsep}{1pt}
      \setlength{\topsep}{3pt}       \setlength{\partopsep}{0pt}
      \setlength{\leftmargin}{1.5em} \setlength{\labelwidth}{1em}
      \setlength{\labelsep}{0.5em} } }
\newcommand{\squishend}{
    \end{list}  }

\newcommand{\hst}{\textit{HST}}
\newcommand{\HST}{\textit{HST}}
\newcommand{\sersic}{S\'{e}rsic}
\newcommand{\JWST}{\textit{JWST}}

\newcommand{\cluster}{SDSS~J1110+6459}
\newcommand{\arcnamelong}{SGAS~J111020.0$+$645950.8}
\newcommand{\arcname}{SGAS~1110}
\newcommand{\zA}{2.481}
\newcommand{\lengthofgal}{7~kpc}
\newcommand{\totalmagnification}{$28\pm8$}
\newcommand{\smallestscale}{$r=30$~pc}  
\newcommand{\rangeofscales}{$r=30$--50~pc}  

\newcommand{\ngalsinhexJ}{$N=3597$}
\newcommand{\ngalsinhexH}{$N=4760$}

\slugcomment{Accepted for publication in The Astrophysical Journal}

\shorttitle{Morphology of a $z=2.48$ star-forming galaxy, with and without lensing}
\shortauthors{Rigby et al.}

\begin{document}

\title{Star Formation at $z=2.481$ in the Lensed Galaxy SDSS J1110$+$6459, II:\\
What is missed at the normal resolution of the \textit{Hubble Space Telescope?}}


\author{
J.~R.~Rigby\altaffilmark{1}, 
T.~L.~Johnson\altaffilmark{2},
K.~Sharon\altaffilmark{2}, 
K.~Whitaker\altaffilmark{3,4,5},
M.~D.~Gladders\altaffilmark{6,7}, 
M.~Florian\altaffilmark{6}, 
J.~Lotz\altaffilmark{8}, 
M.~Bayliss\altaffilmark{9}, \&  
E.~Wuyts\altaffilmark{10}
}
\altaffiltext{1}{Observational Cosmology Lab,
           NASA Goddard Space Flight Center, 8800 Greenbelt Rd., Greenbelt, MD 20771}
\altaffiltext{2}{Department of Astronomy, University of Michigan, 
          500 Church St., Ann Arbor, MI 48109}
\altaffiltext{3}{Department of Astronomy, University of Massachusetts, 710 N Pleasant St., Amherst, MA 01003}
\altaffiltext{4}{Department of Physics, University of Connecticut, 2152 Hillside Road, Unit 3046, Storrs, CT 06269}
\altaffiltext{5}{Hubble Fellow}
\altaffiltext{6}{Department of Astronomy \& Astrophysics, University of
           Chicago, 5640 S. Ellis Ave., Chicago, IL 60637}
\altaffiltext{7}{Kavli Institute for Cosmological Physics, University of
          Chicago, 5640 South Ellis Ave., Chicago, IL 60637}
\altaffiltext{8}{Space Telescope Science Institute, 3700 San Martin Dr., Baltimore, MD 21218}
\altaffiltext{9}{Department of Physics, Massachusetts Institute of Technology}
\altaffiltext{10}{ArmenTeKort, Antwerp, Belgium}

\begin{abstract}
For lensed galaxy \arcnamelong\  at redshift $z=2.481$, 
which is magnified by a factor of \totalmagnification, 
we analyze the morphology 
of star formation as traced by rest-frame ultraviolet emission, in both the 
highly-magnified source plane, and in simulations of how this galaxy 
would appear without  lensing magnification.  
Were this galaxy not lensed but drawn from an \HST\ deep field,
we would conclude that almost all its star formation arises from 
an exponential disk (\sersic\ index of $1.0\pm 0.4$) with an effective radius of 
$r_e = 2.7\pm0.3$~kpc measured from two-dimensional fitting to F606W using Galfit,  
and  $r_e = 1.9 \pm 0.1$~kpc measured by fitting a radial profile to F606W elliptical isophotes.  
At the normal spatial resolution of the deep fields, there is no sign of clumpy star 
formation within \arcnamelong. However, the enhanced spatial resolution enabled 
by gravitational lensing tells a very different story:
much of the star formation arises in two dozen clumps with sizes of 
\rangeofscales\ spread across the \lengthofgal\ length of the galaxy.  
The color and spatial distribution of the diffuse 
component suggests that still smaller clumps are unresolved.  
Despite this clumpy, messy morphology, the 
radial profile 
is still well-characterized by an exponential profile.  
In this lensed galaxy, stars are forming in complexes with 
sizes well below $100$~pc; such sizes are wholly unexplored by 
surveys of galaxy evolution at $1<z<3$.
\end{abstract}


\section{Introduction}

The diffraction limit of the \textit{Hubble Space Telescope} (\textit{HST}), 
$\lambda / D = 0.033$\arcsec\ and 0.052\arcsec\ at  wavelengths of 
3900~\AA\  and 6060~\AA, corresponds to physical scales of 270~pc and 420~pc at 
 $z=2.5$.  By contrast, the typical sizes of the H~II regions that host 
new stars are typically less than $D\sim 100$~pc (c.f.~\citealt{Liu:2013fw}).
As such, diffraction limits our ability to spatially resolve where stars form 
in galaxies in the distant universe.   

Recent work has focused on bright clumps of star formation in distant galaxies 
(e.g. \citealt{Elmegreen:2005fv, Elmegreen:2007id, Elmegreen:2009kd}), 
with typical sizes of $\sim~1$~kpc \citep{ForsterSchreiber:2011by}.
These clumps may arise from gravitational instabilities in gas-rich disks
\citep{Toomre:1964fe, Genzel:2011cp}, or where cold gas has accreted onto the disk
\citep{Keres:2005gb, Dekel:2006cn, Brooks:2009bm}.
 However, the physical scales of these clumps are close to 
the diffraction limit of \HST; it is plausible, even likely, that many of these clumps are 
collections of smaller structures that are blurred together at the spatial resolution 
of \HST\ \citep{Tamburello:2016uk, Fisher:2016jp}.
We have very little evidence as to the presence or importance of star formation on
small ($\la 100$~pc) scales in distant galaxies.
Indeed, this absence of evidence motivates future missions:  
 the science goals for a conceptual $\sim10$~m  ultraviolet and optical space telescope  
include mapping and spectroscopically dissecting star-forming regions in distant galaxies 
down to 100~pc scales \citep{Dalcanton:2015tz}.

Gravitational lensing provides rare opportunities to overcome the \HST\ diffraction 
limit, and study star formation in distant galaxies on small spatial scales
(e.g. \citealt{Swinbank:2009bb, Livermore:2012gw})
Johnson \etal\ (2017a, submitted; hereafter Paper~I) 
presented a detailed lensing model for galaxy cluster~\cluster, 
using a hybrid parametric/non-parametric strong lensing mass model, 
and, use a novel foward-modeling technique to reconstruct in the source plane 
the bright lensed galaxy that it is magnifying, 
\arcnamelong, hereafter \arcname. 
Simulations showed the clump-finding algorithm was $80\%$ complete 
down to an intrinsic clump brightness of $m_{AB}(F606W) = 33.2$, and that the resolution
limit is roughly 20~pc.  Paper I identified two dozen ultraviolet-bright clumps in this 
highly magnified lensed galaxy.
Johnson \etal\ (2017b,  submitted, hereafter Paper~III) 
find that the clump size distribution function is dominated by small clumps
with inferred radii of \rangeofscales .
As such, \arcname\ provides the best opportunity yet to study, at a high spatial
resolution not normally achievable, the morphology of
star formation in a 
galaxy at redshift $z=2.5$.

In this Paper,  we simulate what \arcname\ would look like to \HST\  
were it not lensed, but merely a field galaxy in a deep survey such as 
CANDELS \citep{Grogin:2011hx}.   We analyze its size, structure, 
and morphology at this native unlensed, or 
``candelized'' resolution, and compare to 
the morphology inferred from the reconstructed lensed images.  
Using the test case of \arcname, we explore how inferences about 
star formation in the distant universe depend on the 
available spatial resolution.
Finally, we simulate what \arcname\ would look like, were it not lensed,
 to \JWST\ and to a future large optical space telescope.

We assume a flat cosmology with $\Omega_M = 0.3$, $\Omega_{\Lambda} = 0.7$, 
and $H_0 = 70$~\kms~Mpc$^{-1}$.
In this cosmology, an angular size of 1\arcsec\ 
corresponds to an angular diameter distance of 8.085~kpc at the 
redshift of \arcname\ at z$=$\zA . 

\section{Methods}  

\subsection{Imaging Data}
\begin{figure}
\figurenum{1}
\includegraphics[width=6.5in]{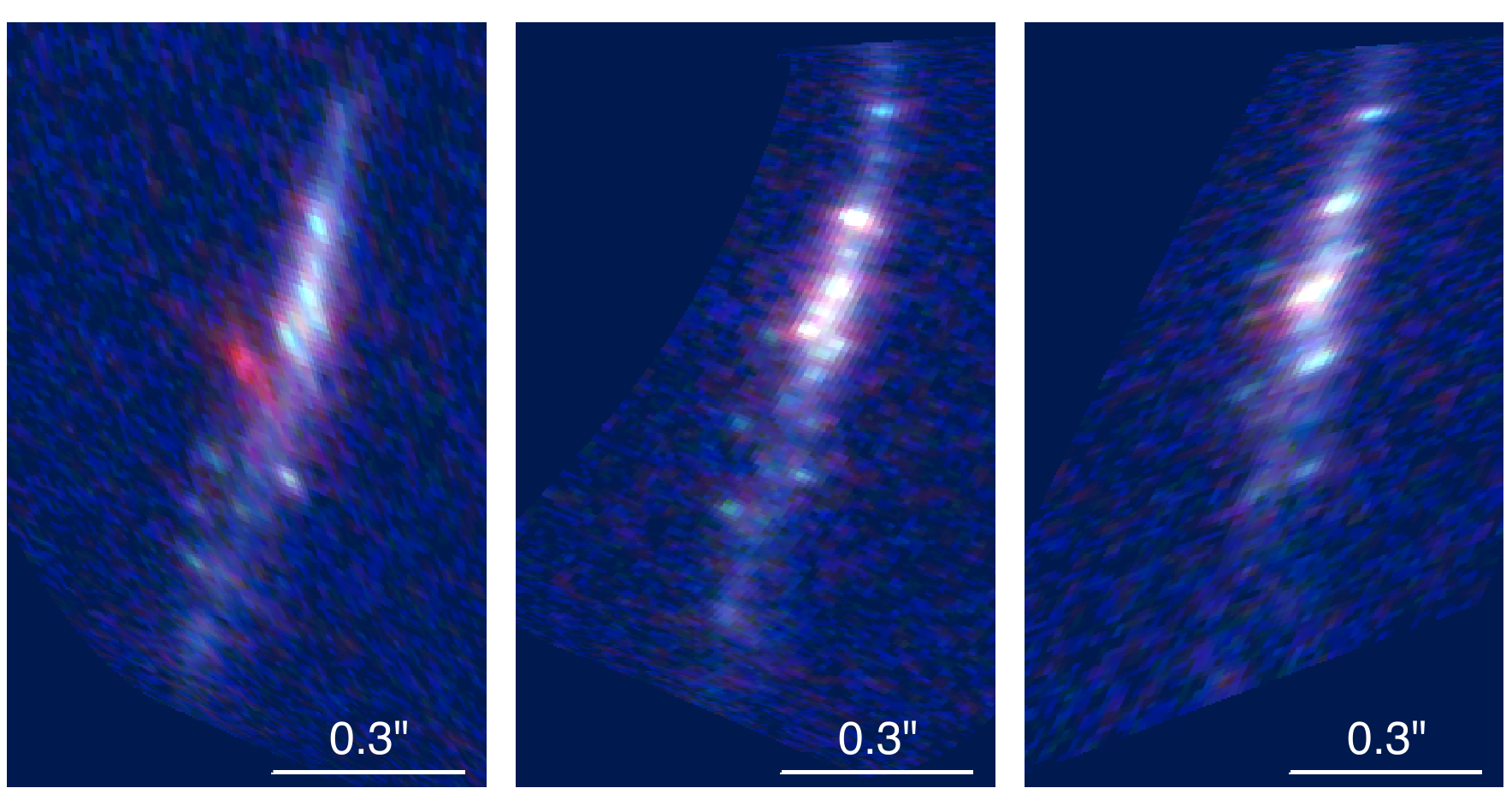}
\figcaption{Source-plane reconstructions of lensed galaxy \arcnamelong\ at z$=$2.481.  
Shown are reconstructions from three images of the lensed galaxy: A1 (left panel), 
A2 (middle panel), and A3 (right panel).  
Image A2 is the most highly magnified and therefore reveals the most detail. 
Image A1 suffers from a contaminating cluster galaxy.
The BGR composite is comprised of F390W, F606W, F105W; the stretch is linear.
 A scalebar of 0.3\arcsec\ is shown.  
}\label{fig:srcplane}
\end{figure}

The data used in this paper are \HST\ WFC3 images in the 
F390W, F606W, F105W, and F160W filters 
from \HST\ program GO~13003 (PI Gladders), and derivative 
source-plane reconstructions as described in Paper~I,
and shown in Figure~\ref{fig:srcplane}.
The data are comparatively shallow: a single orbit in each optical band, 
and half orbit in each infrared band, taken as part of a survey with many targets.

The $5\sigma$ limiting magnitudes (not corrected for lensing) 
quoted in Paper~I in F390W, F606W, F105W, and F160W are
2.5,\footnote{CANDELS did not use the F390W filter; accordingly, 
for CANDELS Deep we quote the depth in the F336W filter.}
5.7, 3.8, and 4.8 mag shallower than the limits quoted  for the CANDELS deep program, 
and --,\footnote{CANDELS Wide did not use the F390W or a similar filter.}
5.1, 3.1, and 4.1 mag shallower than the CANDELS wide program.\footnote{
Using $5\sigma$ magnitude limits from 
http://candels.ucolick.org/survey/Survey\string_Desc.html}
The total magnification of the giant arc \arcname\ is \totalmagnification .
Thus, on average, the intrinsic depth of the \HST\ data for image A2 of \arcname\ 
is comparable to the depth of the CANDELS deep surveys in F105W, 
1 mag deeper in the blue (F390W), and 1.4 to 2 mag shallower in F606W.

Lensing magnification has made the \arcname\ data roughly comparable in 
effective depth to CANDELS; we have not attempted to match depths exactly, 
for example by adding additional noise.  
Instead, as described in the next subsection, we have matched the spatial resolution 
of such observations.

\subsection{Reconstructed images}
We reconstruct all three images of the galaxy \arcname\ in the source plane 
using the methods described in 
\citet{Sharon:2012dr}, \citet{Sharon:2014hh}, and \citet{Sharon:2015hr}, 
by ray-tracing each pixel from the image plane to the source plane.
These are the reconstructions used for most of this paper.
The exception is that when we simulate images from a 
future large ultraviolet and optical telescope 
(\S\ref{sec:jwstluvoir} and Figure~\ref{fig:luvoir}), 
where the highest possible spatial resolution is needed.  For those simulations, 
we use as input the best-fit model of clump positions, sizes, and brightnesses, 
and the best-fit model of the smooth component, in F390W and F606W,
all  in the source plane, from Paper I.   
That model resulted from a forward modeling technique to model the sizes and brightnesses of 
clumps in the source plane, which effectively deconvolves the source galaxy 
from the effects of the lensing PSF.   

\subsection{Convolving to the un-lensed resolution of \HST}
We degraded the source-plane reconstructions  to the  
unlensed resolution of \HST\ as follows.
This procedure was done for each of the three images of the arc: A1, A2, and A3.

For each filter, we used an empirical point spread function (PSF) 
determined from images of 37 lensing clusters 
observed in \HST\ program GO~13003.  The PSFs were created from stars selected 
based on the ratio of flux within a 2\arcsec\ aperture relative to a 0.5\arcsec\ aperture. 
The selected stars were background subtracted and centered, 
$>3\sigma$ background features and nearby objects were masked out, 
and the stars were then averaged 
to generate a high signal-to-noise ratio empirical PSF for each filter. 

For each filter, the empirical PSF was resampled to the same pixel scale
as the source-plane reconstruction, 0.003\arcsec/pix; and then convolved
with the source-plane reconstruction for that filter, using the 
{\bf convolve\_fft} function within the {\bf astropy.convolution} Python package. 

\begin{figure}
\figurenum{2}
\includegraphics[width=6.5in]{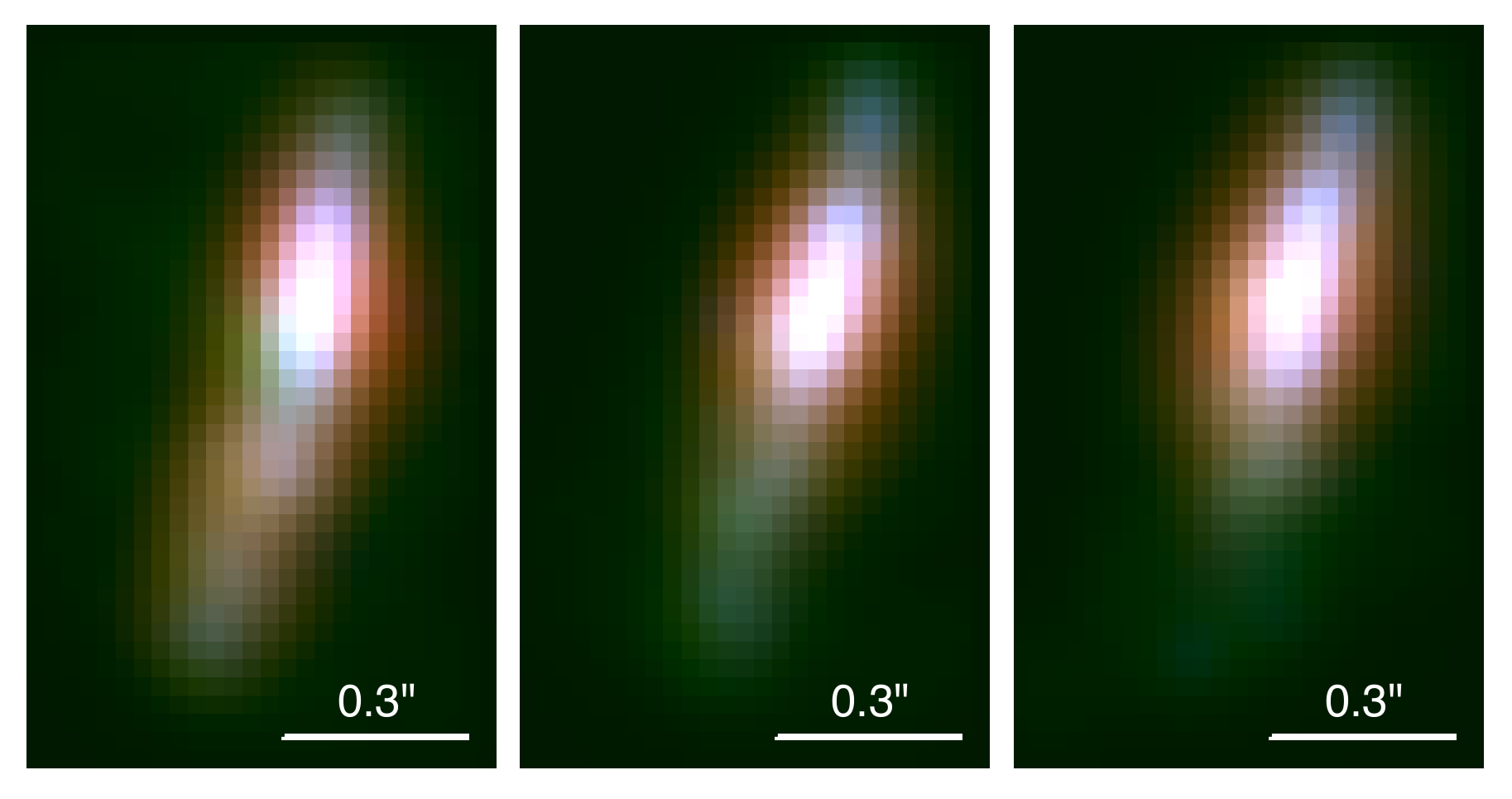}
\figcaption{Simulations of how \arcname\ would appear to \HST\ were it not gravitationally lensed 
but rather a field galaxy.   Shown are the simulated deep field 
images generated from three separate 
images of the lensed galaxy: A1, A2, and A3.
Each filter was  ``candelized'' by convolving the source-plane reconstruction with an empirical 
\HST\ PSF for that filter, and rebinning to a pixel scale of 0.03\arcsec.  
The BGR composite is the same as Figure~\ref{fig:srcplane}. 
A contaminating cluster
galaxy has been subtracted from the F105W of image A1 (left image); 
an artifact of over-subtraction is visible.
}\label{fig:candelized}
\end{figure}

A cluster galaxy contaminates the source-plane reconstruction of image A1.
Using Galfit \citep{Peng:2010eh}, we removed this cluster galaxy 
from the F160W and F105W image-plane
images.  Subtraction residuals can be seen in Figure~\ref{fig:candelized}.

We then rebinned each convolved image to an output pixel scale of 0.03\arcsec/pix. 
These images have been resampled to the normal un-lensed spatial 
resolution of \HST, with a depth approximating  deep surveys like CANDELS.
These ``candelized'' images, or simulated deep field images, are shown in Figure~\ref{fig:candelized}.

\subsection{Convolving to the unlensed resolution of \JWST\ and LUVOIR}\label{sec:jwstluvoir}
Similarly, we simulated the expected spatial resolution of \JWST\ by
convolving the source-plane images with theoretical PSFs 
from a library, version 
``revV-1'',\footnote{\url{http://www.stsci.edu/~mperrin/software/psf_library/}}
generated by Marshall Perrin's WebbPSF tool \citep{Perrin:2014gw}.
These assume the optical error budget from the \JWST\ mission critical design review.
The WFC3 filters F606W, F105W, and F160W were mapped to their closest NIRCam 
equivalents:  F707W, F115W, and F150W. 

We also convolved to the expected spatial resolution of a diffraction--limited large 
space-based ultraviolet/optical/infrared telescope (``LUVOIR''), 
using the F390W source-plane image of \arcname\ A2 as input (since it is
the most highly magnified of the multiple images), and 
scaling the \HST\ F390W empirical PSF by the ratio of the apertures.

These simulated images are shown in Figure~\ref{fig:jwstify} and Figure~\ref{fig:luvoir}.

\subsection{Measuring morphology}
We used the morphology fitting software Galfit \citep{Peng:2010eh}
 to fit a single \sersic\ component to the 
to the simulated unlensed 
data, separately fitting each filter and each lensed image of \arcname. 
For the PSF image required by Galfit, we used the same
empirical PSF used to degrade the resolution.
For the uncertainty required by Galfit, we used a constant uncertainty for each image 
that is the standard deviation of counts.

We fit elliptical isophotes to the source-plane images and the simulated deep field images
using the {\bf Iraf} package {\bf stsdas.analysis.isophote.ellipse}.  
We analyzed the F606W filter because inspection proves it is sensitive to fainter
clumps of rest-frame ultraviolet emission than is F390W; we focused on 
lensed image A2 since it has the
highest magnification of the lensed images of \arcname.  
We use the technique of \citet{Szomoru:2010cd} to account for the PSF;
in short, we add the residuals from the \sersic\ fitting to the best-fit 
\sersic\ model (without the PSF), and fit ellipses to that.

The non-parametric morphology statistics --- Gini coefficient, $M_{20}$, Concentration, and Asymmetry ---  
were measured for the source-plane images and the simulated deep field images 
in each band, using the approach described in 
\citet{Lotz:2004gj, Peth:2016gl}.
The Gini coefficient ($G$, \citealt{Lorenz:1905vb, Abraham:2003dc, Lotz:2004gj})
quantifies the inequality of the light distribution in a galaxy, and is measured using a galaxy's pixels 
with surface brightnesses greater than the surface brightness at its Petrosian radius \citep{Petrosian:1976ea}.
$M_{20}$ is the normalized second-order moment of the brightest regions of the pixels (using the pixel selection as for $G$), 
and quantifies the spatial distribution of bright knots.    
Concentration ($C$; \citealt{Bershady:2000jx, Conselice:2003ds}) is the ratio of the circular radius containing 
$80\%$  ($r_{80}$) of a galaxy's light (as measured within 1.5 Petrosian radii) 
to the radius containing $20\%$ ($r_{20}$) of the light. 
Asymmetry ($A$) is the background-corrected difference between the image of a galaxy and the image rotated by 
180 degrees,  measured with 1.5 Petrosian radii.   
$G-M_{20}$ and $C-A$ have been found to separate galaxies with disturbed or multiply-nucleated morphologies 
from disk and bulge-dominated systems \citep{ Conselice:2003ds, Lotz:2004gj, Lotz:2008kr, Peth:2016gl}.

\section{Results}

\subsection{Star formation rate and stellar mass}
We measure integrated photometry of the lensed images of \arcname, 
and fit spectral energy distribution models, as in \citet{Wuyts:2014eu}, 
to constrain the following {\bf observed} quantities (not corrected for lens magnification):
\begin{itemize}
\item age of 200~Myr (range 100--380); 
\item extinction of $A_v=0.0$ (range 0.0--0.2); 
\item stellar mass of $\log(M_*)=10.68$~\Msun\ (range 10.53--10.79);
\item  star formation rate of 230 \Msun yr$^{-1}$   (range 220--440).
\end{itemize}

To translate into {\bf intrinsic} quantities, we compute the appropriate magnification, by
taking the ratio of the area of the photometry aperture in the image plane to the ray-traced
area of that aperture in the source plane.  We compute this magnification for each of the 
eight lens models in Paper~I;  the median and median absolute deviation
of the magnifications is \totalmagnification.

As such, we estimate the {\bf intrinsic} quantities as:
\begin{itemize}
\item stellar mass of $\log M^* = 9.24$~\Msol, with associated uncertainties of 
 $^{+0.11}_{-0.15}$  from SED fitting and 
$^{+0.08}_{-0.12}$ from the magnification uncertainty.
\item star formation rate as $8.5$~\Msol yr$^{-1}$, with associated uncertainties of 
$^{+8}_{-0.4}$ from SED fitting and 
$^{+4}_{-2}$ from the magnification uncertainty.
\end{itemize}

\subsection{Clumpiness and color}
The discrete rest-frame ultraviolet clumps identified in Paper~I 
contain a total of  
23\%  (22\%)   of the light in the F390W (F606W) filter in the image plane. 
Formally, these measurements are lower limits on the percent of the light in clumps; 
there are doubtless clumps too faint or too small for our data to identify, that we 
have lumped into the ``smooth'' component.

We measure the rest-frame ultraviolet color of the clumps and the diffuse emission, 
at the full spatial resolution offered by gravitational lensing. 
The mean  color and standard deviation of the clumps is 
$F390W - F606W = 0.48 \pm 0.39$, measured in the source plane.
For the smooth component, we measure the flux-weighted color in the image plane to be 
$F390W - F606W = 0.45 \pm 0.05$, with negligible variation 
in color depending on the surface brightness cut adopted.
Thus, the clumpy and diffuse emission have the same average 
color within uncertainties, 
implying similar stellar populations and reddening.  
The spatial distribution of the smooth component also closely traces the 
spatial distribution of the clumps; see Figure~9 of Paper~I.

\subsection{Size and structure}
\label{sec:results_size}
At the full spatial resolution, the majority of the rest-frame ultraviolet light 
(52\% percent, measured in either F390W or F606W) in \arcname\
is concentrated within the central 0.3\arcsec\ (2.4~kpc).  

At the normal resolution of \HST\ deep fields, 
an extended central component contains most of the flux:
the single \sersic\ component fit by Galfit that
best fits each image  (averaging over all 4 bands and all 3 images)
 has a \sersic\ index of $1.0 \pm 0.4$, and   
an effective radius of $2.3\pm 0.3$~kpc; 
that component contains $97\% \pm 7\%$ of the total light.  
Table~\ref{tab:galfitresults} lists these results for each filter and each image 
of the lensed arc; Figure~\ref{fig:galfit} shows moderate trends with 
wavelength, which are unsurprising given the factor of three difference in 
diffraction--limited resolution between the bluest (F390W) and the 
reddest (F160W) filters.

We now contextualize the physical sizes we measure with Galfit for the simulated deep field
images of \arcname, with measurements and samples from the literature.  We first consider
the older stellar population, using the F160W filter. 
\citet{vanderWel:2012eu}  fit Galfit profiles to a large sample of unlensed galaxies 
from the CANDELS survey.  
From the catalogs of  \citet{vanderWel:2012eu} and \citet{Momcheva:2015wa}, 
we select a  subset of galaxies at matched
redshift and stellar mass to \arcname.  
These have  an acceptable Galfit fit (flag$=$0) in the F160W filter, 
a 3D-HST ``best redshift`` of $2<z<3$, 
and a stellar mass in the range $9.0<\log M^*<9.5$~\Msol.
A total of \ngalsinhexH\ galaxies satisfy these criteria.
Figure~\ref{fig:vanderwel} compares the results for this matched sample to 
those of \arcname.
The \sersic\ index we measure for the  simulated deep field
\arcname\ images are entirely consistent with the average \sersic\ index 
from \citet{vanderWel:2012eu}. 
This CANDELS subset has $R_e = 1.6 \pm 0.6$~kpc
(median $\pm$ median absolute deviation) for F160W,  
compared to $R_e = 2.6 \pm 0.2$~kpc for \arcname\ in F160W.  
Thus, in F160W, \arcname\ has an effective radius that is larger than average, 
but well within the observed range, of galaxies at matched stellar mass and redshift.

We examine the sensitivity of this result to age, by dividing the 3D-HST galaxies into 
two groups, with specific star formation rate (as listed in the 3D-HST catalog) above or 
below 1~$Gyr^{-1}$.  The older and younger subsets have, respectively, effective radii 
in F160W of $R_e = 1.55 \pm 0.64$~kpc and  $R_e = 1.70 \pm 0.64$~kpc
(median and median absolute deviation). 
Thus, the galaxies with younger ages have somewhat larger sizes.

We repeat this analysis for the filter F125W.
In F125W,  a total of \ngalsinhexJ\ galaxies in 
\citet{vanderWel:2012eu} satisfy the redshift and stellar mass ranges.
These have a median effective radius of 
$R_e = 1.7 \pm 0.7$~kpc  (median $\pm$ median absolute deviation).  
The older and younger subsamples, respectively, have 
 $R_e = 1.66 \pm 0.68$~kpc and  $R_e = 1.74 \pm 0.65$~kpc.  
These results are consistent with those for F160W.

Figure~\ref{fig:profiles} plots the radial profiles of 
F606W surface brightness that resulted from 
fitting elliptical isophotes to lensed image A2.  
The isophotes of both the source-plane image and the simulated deep field 
image are fit well by exponential profiles. 
For an exponential profile, the disk scale length $r_s$ is related to the effective radius $r_e$ as: 
$r_e = 1.678~r_s$.  
Converting, the effective radii from the elliptical isophote fitting are 
$r_e = 1.5 \pm 0.1$~kpc for the source-plane image and $r_e = 1.9 \pm 0.1$~kpc for the 
simulated deep field image, both in F606W.

These sizes are somewhat smaller than the typical effective radii of 
$r_e = 2.5$~kpc 
measured by \citep{Elmegreen:2005ipa} for 
for disk galaxies in the \HST ultra-deep field measured with 
the ACS F775W filter.

Comparing these multiple measurements,
for the elliptical isophoe fitting, larger sizes are measured
from the simulated deep fields than from the source-plane reconstructions;
this is presumably a resolution effect.
For the simulated deep fields, the two-dimensional Galfit fitting returns a
a sizes that is 1.3 $\sigma$ larger than derived from elliptical isophote fitting.

\subsection{Quantitative morphology}
In Figure~\ref{fig:lotz}, we compare the quantitative morphological measurements 
of Gini, M20, Concentration, and Asymmetry of \arcname\ to those measured for 
CANDELS galaxies matched in redshift and stellar mass  
($2<z<3$, $9.0<\log ( M_*/M_{\odot} <9.5$). 
We consider the filters F606W, F105W, and F160W, and measure at both the source plane resolution 
and at the simulated deep field resolution.  We find negligible difference between $C$, $G$, and $M_{20}$
values for the source-plane and simulated deep field versions of the \arcname\ measures.   
The measured Asymmetry values for the source-plane images are unphysically negative,  
likely due to an overcorrection for the asymmetry of the background. 

\arcname\ would likely be classified as a ``Group 1 galaxy'' \citep{Peth:2016gl},
meaning it has a late-type disk without a prominent bulge.  
Its asymmetry is insufficiently high to be considered a merger;  
neither $G$ nor $M20$ is high,  indicating that \arcname\ does not have any particularly prominent clumps.    
Compared to the morphology distributions of matched CANDELS galaxy sample,  
SGAS is a typical disk galaxy for its redshift and stellar mass.

\section{Discussion}\label{sec:discussion}
\arcname\ is a rare case where high lensing magnification provides 
a much sharper view of a distant galaxy than is normally possible.
It is therefore appropriate to take the lensing reconstruction as the ``truth image'', 
and consider to what extent  deep, non-lensed \HST\ images could 
recover that morphology. 

In short, at the normal spatial resolution of \HST, \arcname\ is 
correctly classified as a non-merging  disk galaxy; its 
size is correctly measured, as is the fact that its light profile is exponential. 
What is missed, however, is extreme clumpiness of the star formation 
on $<100$~pc scales.

Were \arcname\ not gravitationally lensed but instead drawn from 
an \HST\ deep field,  we would conclude that 
almost all of the star formation emerges from an exponential 
disk (\sersic\ index of $1.2\pm 0.3$ in F606W) with an effective radius of 
$r_e = 2.7\pm0.3$~kpc measured from two-dimensional fitting to F606W using Galfit,  
and $r_e = 1.9 \pm 0.1$~kpc measured from 1D fits to the elliptical isophotes,
after correcting for the PSF.
For its stellar mass and redshift, \arcname\ at F160W 
($\lambda_{rest} \sim 0.65$~\micron) 
appears larger than average, but well within the observed range.  
In the rest-frame ultraviolet 
as probed by the F390W and F606W filters ($\lambda_{rest} \sim 0.1$--0.3~\micron),
we were unable to find a matched sample, 
but the size of \arcname\ appears typical of those measured
for star-forming galaxies at lower or similar redshift.  

Quantitative morphological measures would classify \arcname\ as a disk galaxy, without signs of
a merger or prominent clumps.  The non-parametric measures $G$, $M_{20}$, $C$ are consistent
with CANDELS galaxies at this epoch and mass scale.  The rest-frame $\sim$ 4600\AA\ morphology values
(as probed by F160W) are comparable to those measured for local late-type Sc/Sd/dIrr galaxies (e.g. \citealt{Lotz:2004gj}.) 
Interestingly, the quantitative morphological measurements are consistent for the 
simulated deep field  resolution and
the full lensed resolution.   This suggests that the robustness of these measures,
which has been demonstrated for spatial resolution between
$\sim 100$ pc and $\sim$ 1 kpc in local galaxies  (see Figure 6 of \citealt{Lotz:2004gj}) also holds for
higher redshift galaxies.

At the simulated deep field resolution, \arcname\ shows no obvious off-nuclear 
``blobs'' or ``clumps'' of star formation that contain more than $8\%$ of 
the total UV luminosity, which is the clump definition proposed by \citet{Guo:2015dr}.  

Thus, were \arcname\ in an \HST\ deep field, one would 
conclude that it is an inclined galaxy undergoing 
centrally--concentrated star formation in a smooth, 
$r_e$=2~kpc exponential disk, with no off-axis clumps of star formation.
 How does this picture compare to that revealed 
by the source plane images at full spatial resolution?

The reconstructed source-plane images, which resolve clumps with radii down to 
\smallestscale\  (Paper I), show that \arcname\ is forming stars
across its \lengthofgal\ length.  
Half the rest-frame ultraviolet light 
(52\% percent, measured in either F390W or F606W) 
is concentrated within the central 0.3\arcsec\ (2.4~kpc). 
A significant minority of the rest-frame ultraviolet light, 
23\% at F390W and 22\% at F606W, resolves into more than twenty 
discrete clumps identified in Paper~I.

The smooth component and the clumps have very similar spatial distribution, 
and identical rest-frame ultraviolet color within uncertainties.  
This implies that the smooth component and the clumps have similar
star formation histories and extinction.   Much of that ``smooth'' component 
may be comprised of smaller star-forming regions  that are still unresolved.  
Some of that emission could also be truly diffuse, arising from stars
that escaped from short-lived star clusters, or from genuine ``field'' stars
that were born outside star clusters \citep{Massey:1995gk}.

Despite the extremely clumpy morphology of the star formation, the
elliptical isophotes fit to the F606W source-plane reconstruction are 
well-described by an exponential profile. 
While the two-dimensional morphology is complex, it averages
out to a smooth one-dimensional surface brightness profile.
As such, our results extend, down to much smaller physical scales, 
 a result of \citet{Elmegreen:2005ipa}, that clumps in 
UDF spiral and irregular disk galaxies follow an exponential distribution of 
luminosity versus radius.  
Those authors suggest that these clumps tidally disperse to form a disk; 
\citet{Genzel:2008dt} suggest they may migrate inward to build up bulges or thick disks.
This result also bears on the smooth, large-scale formation of stars in 
exponential disks traced in H$\alpha$ in distant galaxies by \citet{Nelson:2012kb}.  
Our results suggest that such star formation, which appears smooth on kiloparsec scales 
and when averaged over dozens of galaxies, 
is in fact highly clumpy and messy on smaller ($\la$100) parsec scales in 
individual galaxies.

Given the complexity of local galaxies, where stars form in star-forming regions ranging from 
parsec-scale single--star H~II regions up to $\ga 100$~pc complexes like 30~Doradus 
and Carina, it may not be surprising that star formation in the distant universe can 
have significant structure on $<100$~pc scales.  \arcname\ is the best
demonstration to date that such processes are also at work in the distant universe.

\begin{figure}
\figurenum{3}
\includegraphics[width=6in]{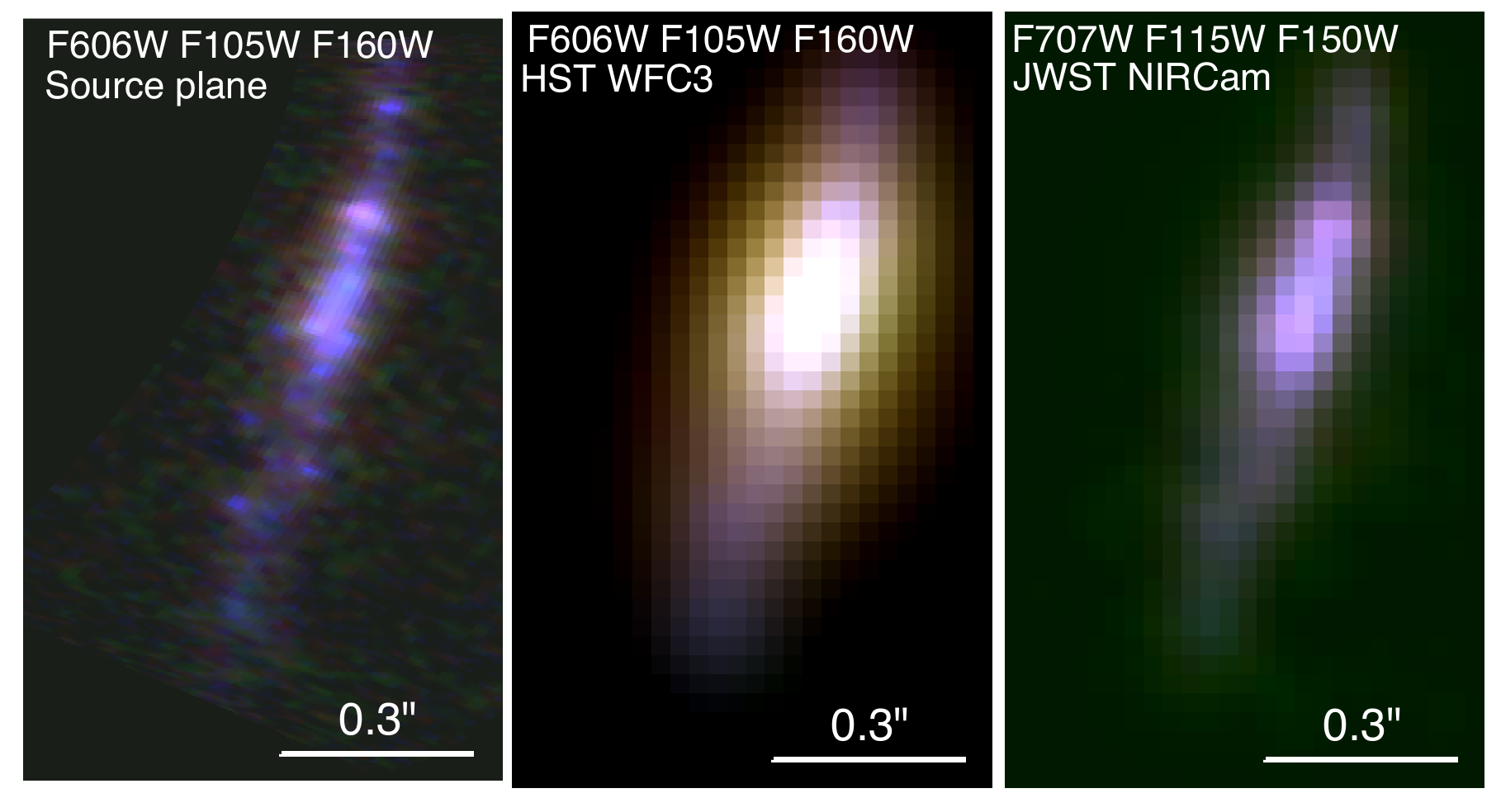}
\figcaption{Simulation of how \arcname\ without lensing 
would look to WFC3/\textit{HST} and  NIRCam/\textit{JWST}.
{\bf Left panel:}  \HST\ WFC3 source-plane image in F606W, F105W, and F160W;
{\bf Middle panel:} \HST\ WFC3 simulated without lensing in F606W, F105W, and F160W;
{\bf Right panel}:  \JWST\ NIRCam simulated without lensing in F707W, F115W, and F150W.
Lensed image A2 was the input for these simulations.  
}\label{fig:jwstify}
\end{figure}

\begin{figure}
\figurenum{4}
\includegraphics[width=6.5in]{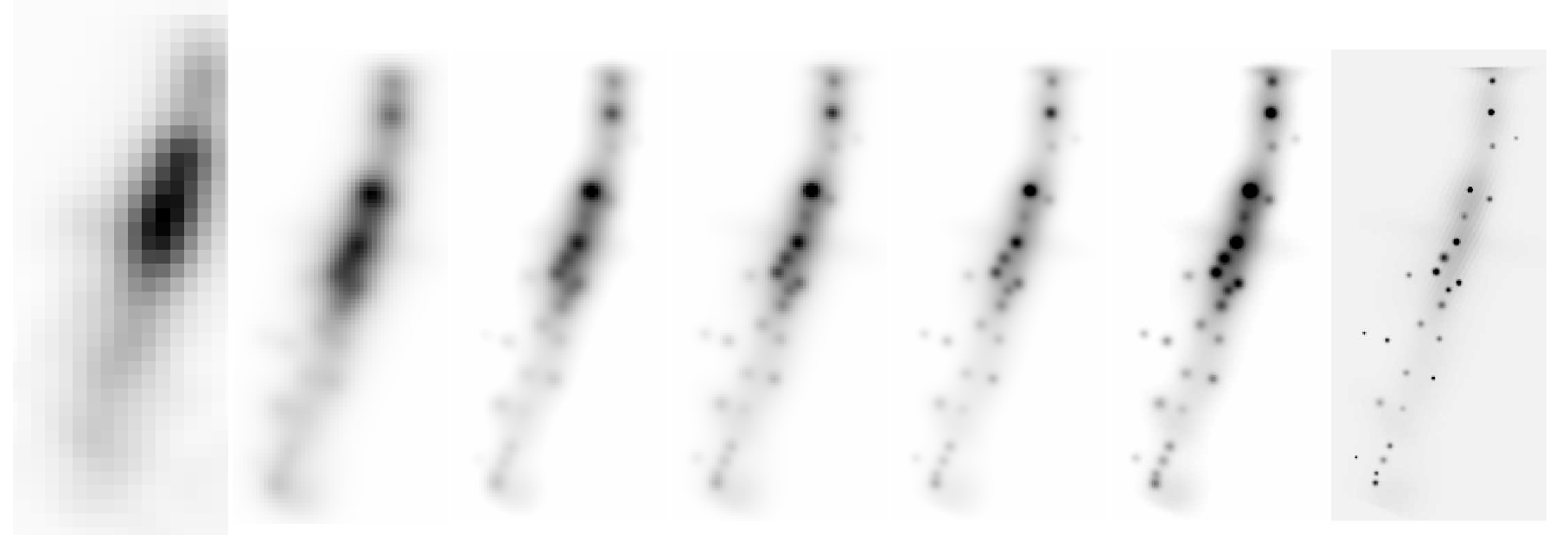}
\includegraphics[width=6.5in]{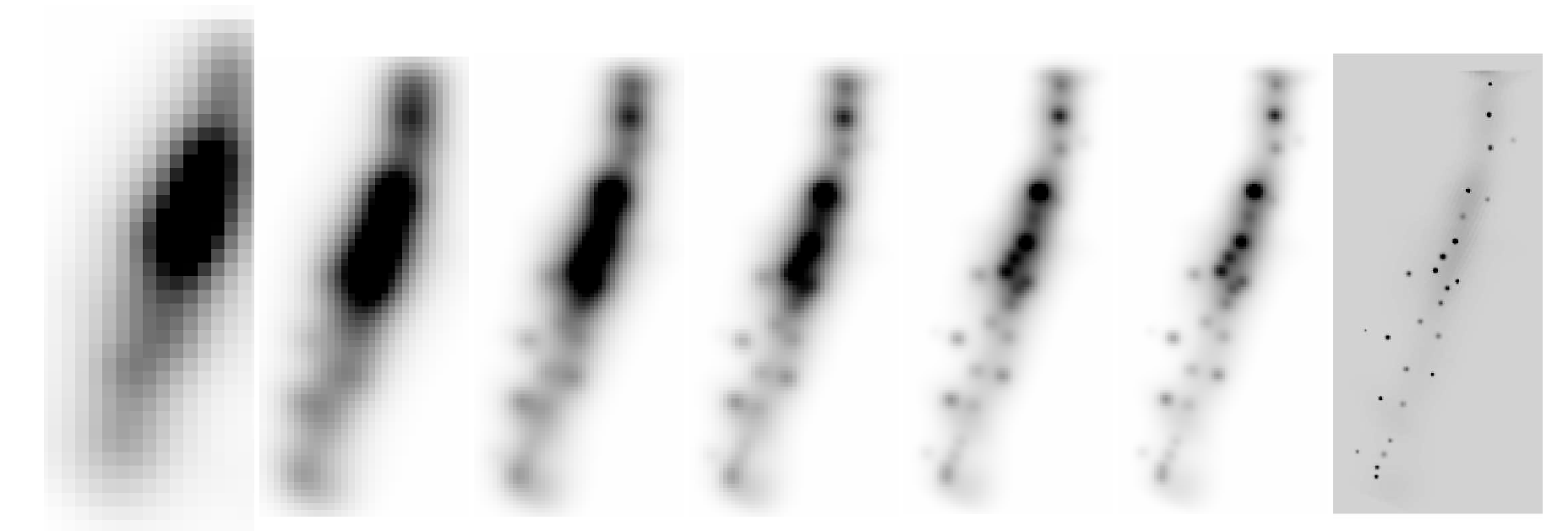}
\figcaption{Simulation of how \arcname\ without lensing 
would look to a diffraction-limited large ultraviolet/optical/infrared telescope (``LUVOIR'') of varying aperture size.
The PSF has been scaled from the empirical \HST\ PSF, and the binning is Nyquist sampled. 
The top panel is F390W, and the bottom panel is F606W.
From left to right, the columns are:  HST (2.4m), 4m, 6m, 8m, 10m, 12m, and the noise-free model of 
the  lensed source-plane reconstruction, from which the other images were derived.
Lensed image A2 was the input for these simulations.  No noise has been added.
}\label{fig:luvoir}
\end{figure}

\begin{figure}
\figurenum{5}
\includegraphics[width=5in]{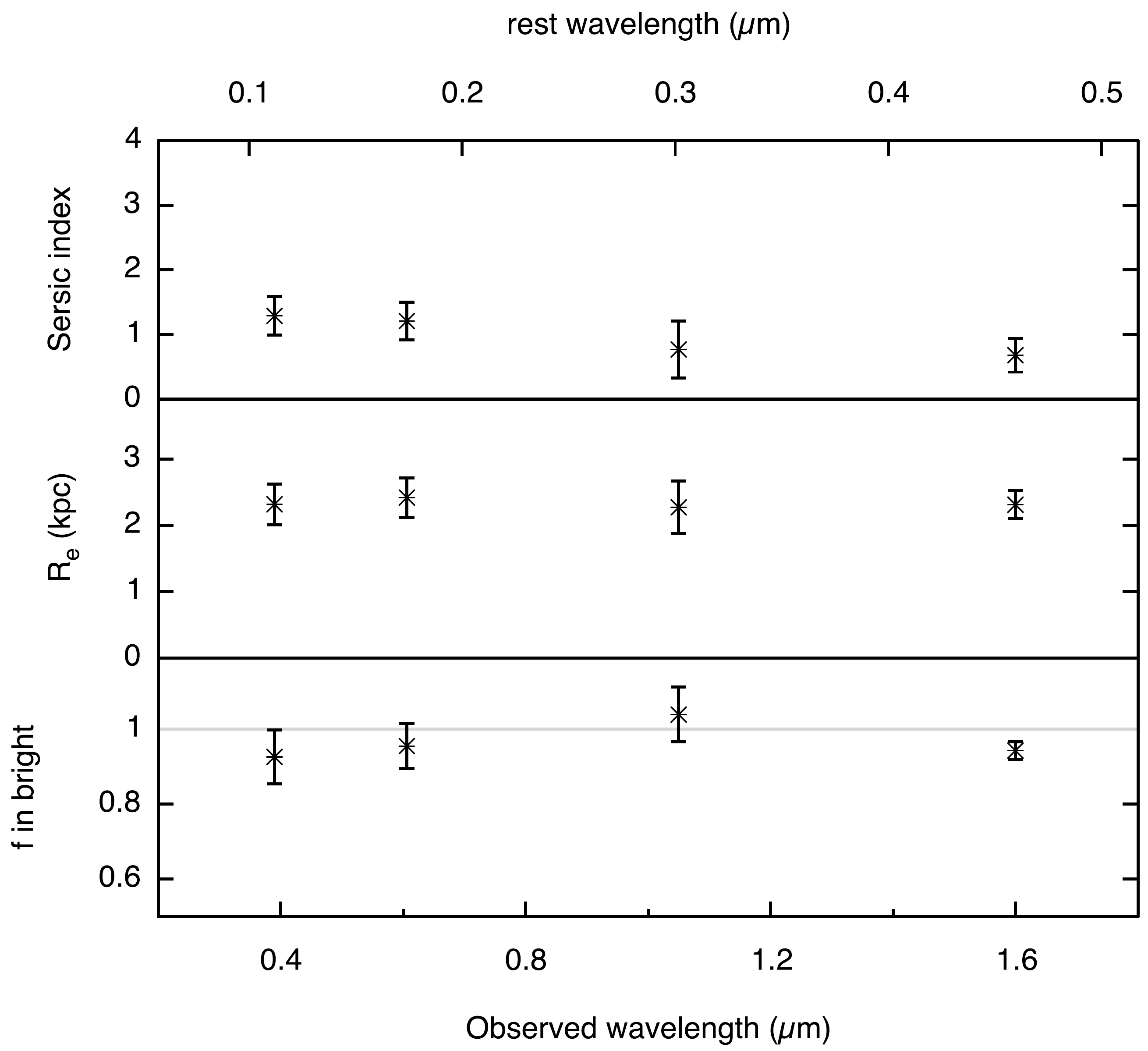}
\figcaption{Results of fitting a single bright component to each simulated deep field 
image of \arcname.
Plotted are the \sersic\ index,  the effective radius, 
and the fraction of the total light fit by that single component.
Measurements were made separately for each of the three lensed images of \arcname; 
for each filter we plot the average of the measurements  and the standard deviation.  
These measurements are tabulated in Table~\ref{tab:galfitresults}.
}\label{fig:galfit}
\end{figure}

\begin{figure}
\figurenum{6}
\includegraphics[width=4in]{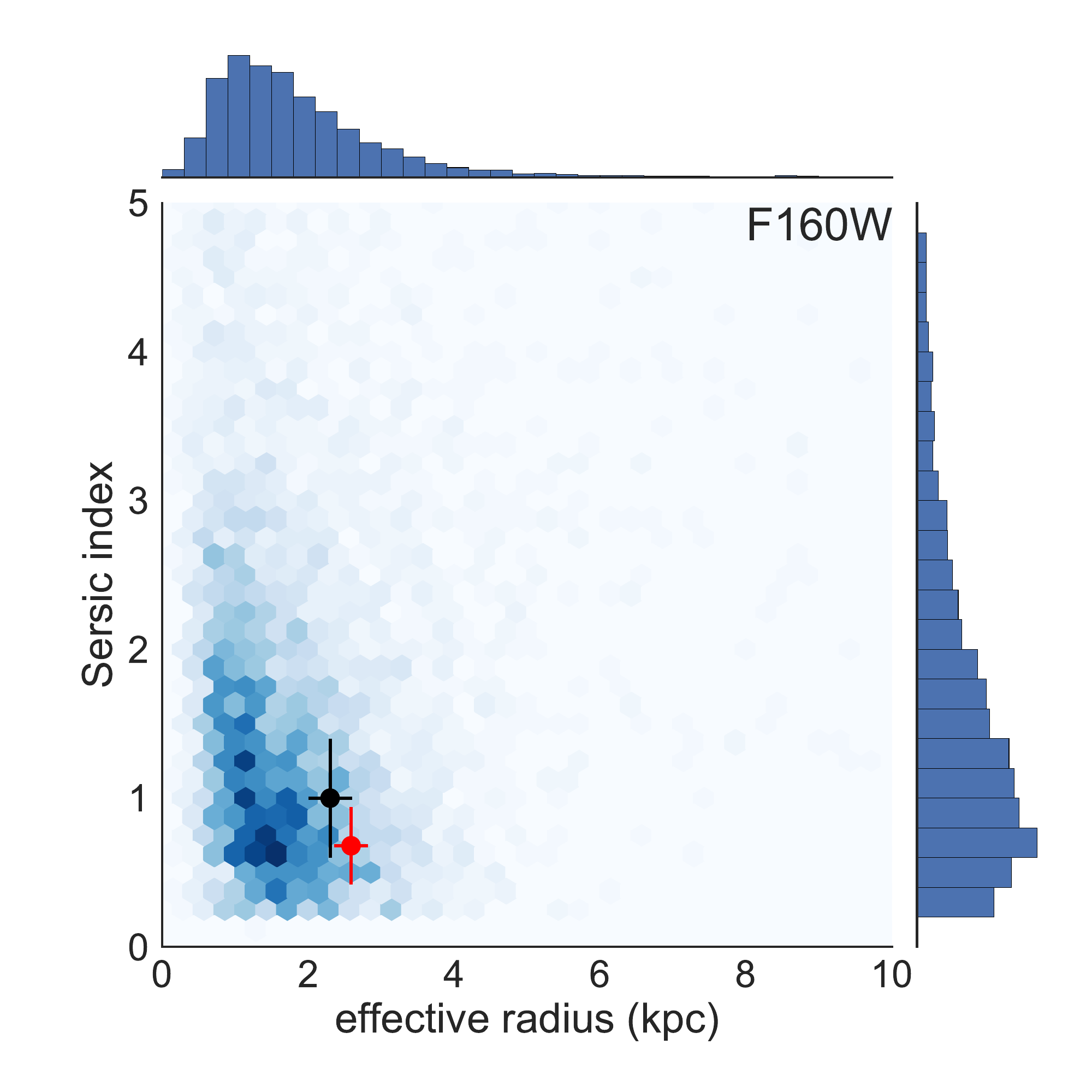}
\figcaption{Comparison of \sersic\ indices and effective radii to CANDELS. 
A density map shows measurements for the \ngalsinhexH\ galaxies in the 
catalogs of \citet{vanderWel:2012eu} that have good Galfit fits (``FLAG=0'') in the F160W filter, 
3D-HST ``best'' redshift of $2<z<3$, and 
stellar masses of $9.0<log M^* < 9.5$~\Msol\ from 3D-HST \citep{Momcheva:2015wa}.  
The red point shows our measurement for the simulated deep field  F160W image of \arcname, 
using Galfit, from \S\ref{sec:results_size};
the black point shows the average measurements for the simulated deep field 
images in all four bands. 
Histograms of effective radius and \sersic\ index are shown in the margins. 
In F160W, the  CANDELS subset has $R_e = 1.7 \pm 0.7$ kpc 
(median $\pm$ median absolute deviation), 
compared to $R_e = 2.6 \pm 0.2$ kpc  for \arcname\ in that filter.
}\label{fig:vanderwel}
\end{figure}

\begin{figure}
\figurenum{7}
\includegraphics[width=6.5in]{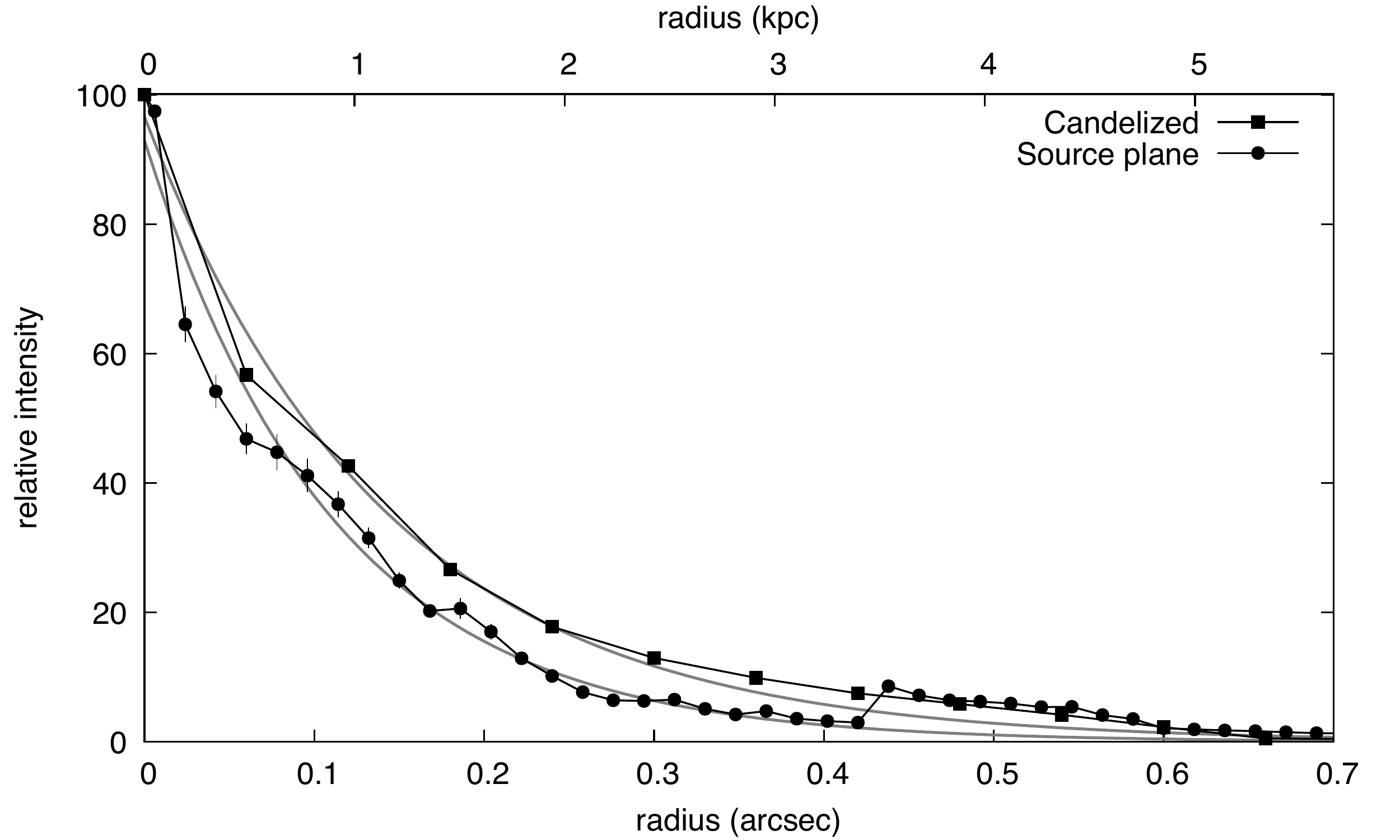}
\figcaption{Distribution of intensity with radius, from isophotal 
ellipses fit to the source plane F606W reconstruction of A2 (filled circles)  and the 
simulated deep field F606W for 
A2 (filled squares), using 
the method of \citet{Szomoru:2010cd}  to account for the PSF.
The best-fitting exponential profiles are overplotted in grey.  
The profiles have been scaled to a relative intensity of $100$ at $R=0$.
}\label{fig:profiles}
\end{figure}

\begin{figure}
\figurenum{8}
\includegraphics[width=6in]{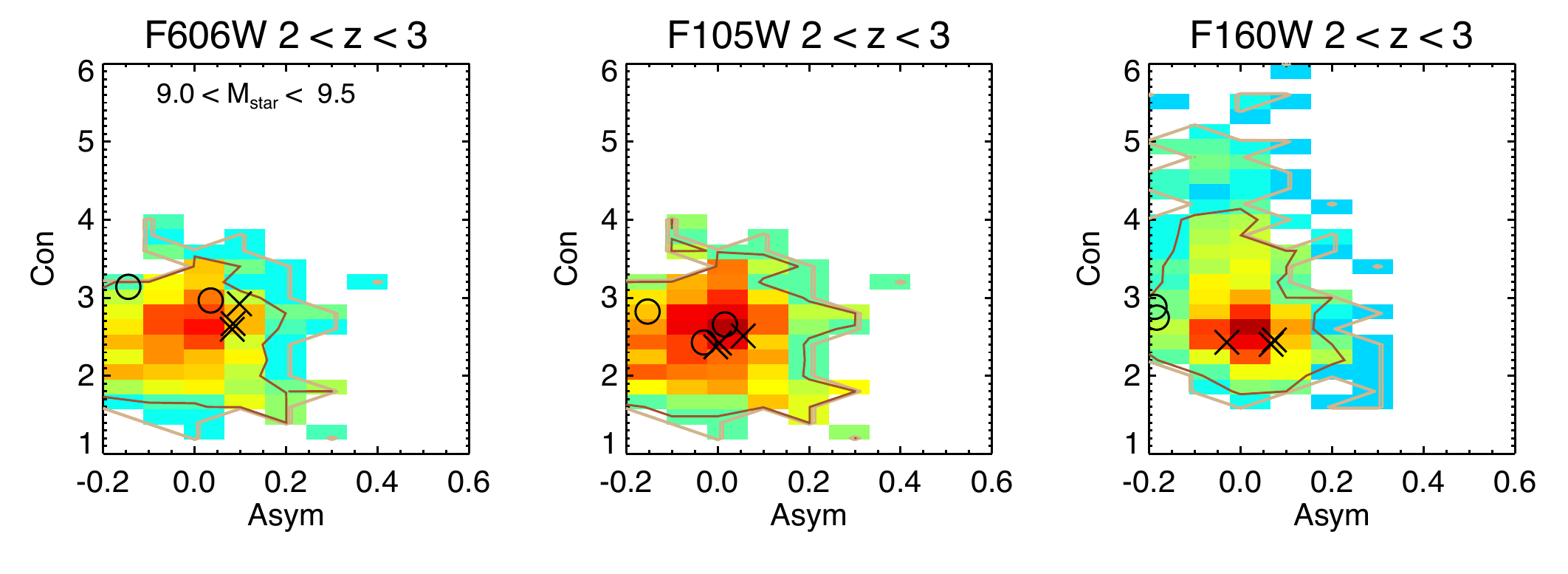}
\includegraphics[width=6in]{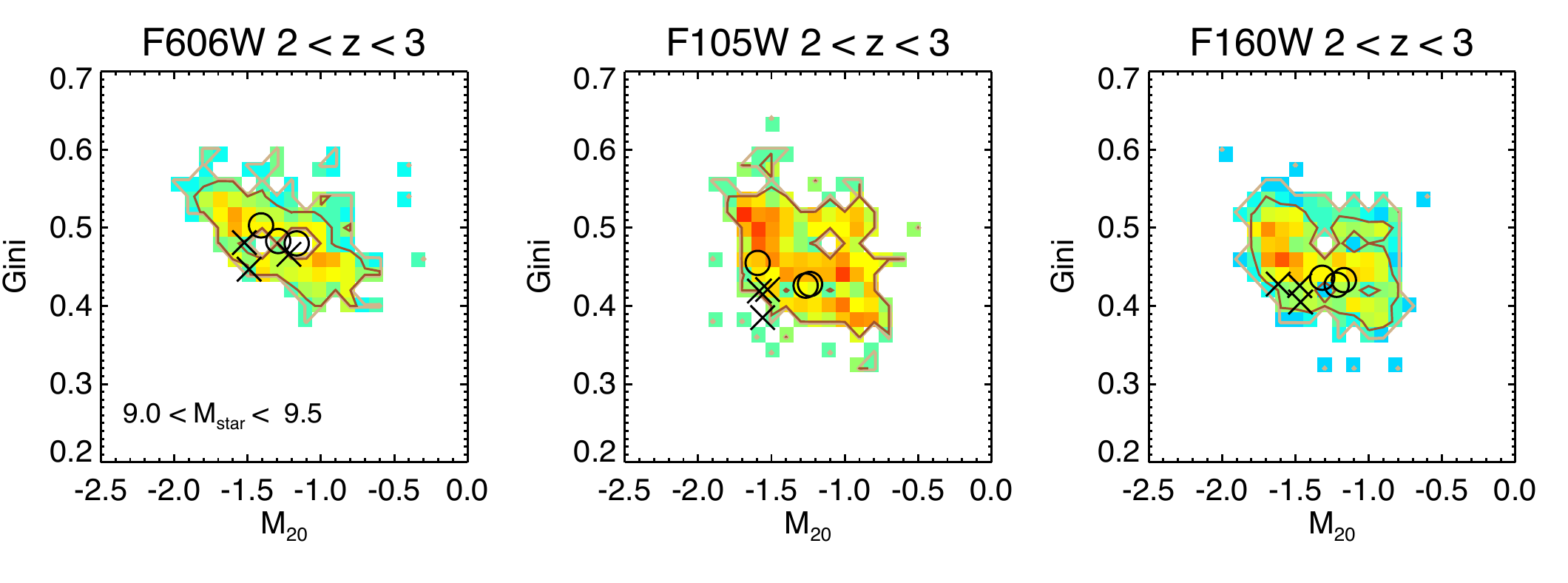}
\figcaption{Quantitative morphological measurements.  Measurements for \arcname\
are shown at the resolution of the source-plane reconstruction \textit{(circles)}, and 
at the simulated deep field resolution  \textit{(crosses)}.  The background contours and color 
are measurements for the subset of CANDELS galaxies with redshift in the range 
$2<z<3$ and  stellar mass in the range $9.0 < \log (M_* /M_{\odot}) < 9.5$. 
}\label{fig:lotz}
\end{figure}

Looking toward the future, our results suggest that 
surveys of galaxy morphology at $z\sim1$--3 have not yet surveyed the 
critical size scales of star-forming regions.  
Figure~\ref{fig:jwstify} shows that \JWST\ will spatially resolve some of those size scales. 
However, spatially resolving the dozens of star-forming regions visible in the 
source-plane image of \arcname\ would require a much larger telescope, for example
a 10~m aperture working in observed blue optical (rest-frame ultraviolet).
The importance of star formation on such small physical scales at $z\sim2$ should
inform the mission concepts for future large telescopes to probe distant galaxies
in the rest-frame ultraviolet.   Our results imply 
that there is significant sub-kiloparsec structure for large future telescopes
to explore with imaging and spectroscopy.

\acknowledgments
This paper is based on observations made with the 
NASA/ESA Hubble Space Telescope, obtained at the Space Telescope Science Institute, 
which is operated by the Association of Universities for Research in Astronomy, Inc., 
under NASA contract NAS 5-26555. 
These observations are associated with \hst\ program \# 13003.
Support for \hst\  program \# 13003 was provided by NASA through a grant from
the Space Telescope Science Institute, which is operated by the
Association of Universities for Research in Astronomy, Inc., under
NASA contract NAS 5-26555. 
TLJ acknowledges support by NASA under Grant Number NNX16AH48G.
KEW acknowledges support by NASA through Hubble Fellowship grant
\#HF2-51368 awarded by the Space Telescope Science Institute, which is
operated by the Association of Universities for Research in Astronomy,
Inc., for NASA, under contract NAS 5-26555.
This research has made use of open-source Python
packages including SciPy (\url{http://www.scipy.org/}), 
NumPy \citep{vanderWalt:dp}, 
IPython \citep{Perez:2007hy},
Pandas \citep{McKinney:2010un}, 
and AstroPy \citep{TheAstropyCollaboration:2013cd}.
JRR thanks the organizers of the Python Bootcamp at the 
NASA Goddard Space Flight Center.  
JRR acknowledges the hospitality of the Astronomy department at 
Michigan State University, where this paper was begun during a blizzard.
JRR is grateful to the late Fred Lo for useful discussion and encouragement.
{\it Facilities:} \facility{HST (WFC3)}

\bibliographystyle{astroads}
\bibliography{papers}
\clearpage
\begin{deluxetable}{llllllllllllllll}
\tabletypesize{\small}
\tablecolumns{16}
\tablewidth{0pc}
\tablenum{1}
\tablecaption{Galfit results, fitting a \sersic\ component to each candelized image\label{tab:galfitresults}}
\tablehead{
&  \multicolumn{5}{l}{Sersic indices} & \multicolumn{5}{l}{Effective radius $r_e$ (\arcsec)} & \multicolumn{5}{l}{Fraction of light in model}\\
\colhead{filter} &  \colhead{A1} &  \colhead{A2} &  \colhead{A3} &  \colhead{mean} &  \colhead{std} &  
\colhead{A1} &  \colhead{A2} &  \colhead{A3} &  \colhead{mean} &  \colhead{std} &
\colhead{A1} &  \colhead{A2} &  \colhead{A3} &  \colhead{mean} &  \colhead{std}
}
\startdata
F390W  & $1.44$ & $1.50$ & $0.95$ & $1.29$ & $0.30$ & $0.35$ & $0.34$ & $0.27$ & $0.32$ & $0.043$ & $0.86$ & $1.00$ & $0.92$ & $0.93$ & $0.07$  \\
F606W  & $1.10$ & $1.54$ & $0.99$ & $1.21$ & $0.29$ & $0.35$ & $0.36$ & $0.29$ & $0.33$ & $0.041$ & $0.89$ & $1.00$ & $0.97$ & $0.96$ & $0.06$  \\
F105W  & $0.36$ & $1.23$ & $0.70$ & $0.77$ & $0.44$ & $0.37$ & $0.31$ & $0.26$ & $0.31$ & $0.055$ & $1.12$ & $1.00$ & $0.99$ & $1.04$ & $0.07$  \\
F160W  & $0.38$ & $0.84$ & $0.83$ & $0.68$ & $0.26$ & $0.35$ & $0.31$ & $0.29$ & $0.32$ & $0.029$ & $0.92$ & $0.95$ & $0.96$ & $0.94$ & $0.02$  \\
\enddata
\tablecomments{Effective radii are quoted in arcseconds; the pixel scale was 0.03\arcsec per pixel.}
\end{deluxetable}

\clearpage

\end{document}